\documentclass[preprint2]{aastex62}
\usepackage{amsmath}
\begin{document}

\title{On the role of a cavity in the hypernova ejecta of GRB 190114C}

\correspondingauthor{ \\R. Ruffini, J. D. Melon Fuksman, G. V. Vereshchagin}
\email{ruffini@icra.it, \\dmelonf@gmail.com, \\veresh@icra.it}

\author{R. Ruffini}
\affiliation{ICRANet, P.zza della Repubblica 10, 65122 Pescara, Italy}
\affiliation{ICRA and Dipartimento di Fisica, Sapienza Universit\`a di Roma, P.le Aldo Moro 5, 00185 Rome, Italy}
\affiliation{INAF, Viale del Parco Mellini 84, 00136 Rome, Italy}

\author{J. D. Melon Fuksman}
\affiliation{ICRANet, P.zza della Repubblica 10, 65122 Pescara, Italy}
\affiliation{ICRA and Dipartimento di Fisica, Sapienza Universit\`a di Roma, P.le Aldo Moro 5, 00185 Rome, Italy}

\author{G. V. Vereshchagin}
\affiliation{ICRANet, P.zza della Repubblica 10, 65122 Pescara, Italy}
\affiliation{INAF -- Istituto di Astrofisica e Planetologia Spaziali, 00133 Via del Fosso del Cavaliere, 100, Rome, Italy.}
\affiliation{ICRANet-Minsk, B.I. Stepanov Institute of Physics, National Academy of
Sciences of Belarus, 220072 Nezalezhnasci Av. 68-2, Minsk, Belarus}

\begin{abstract}
Within the binary-driven hypernova I (BdHN I) scenario, the gamma-ray burst GRB190114C originates in a binary system composed of a massive carbon-oxygen core (CO$_{core}$), and a binary neutron star (NS) companion. As the CO$_{core}$ undergoes a supernova explosion with the creation of a new neutron star ($\nu$NS), hypercritical accretion occurs onto the companion binary neutron star until it exceeds the critical mass for gravitational collapse. The formation of a black hole (BH) captures $10^{57}$ baryons by enclosing them within its horizon, and thus a cavity of approximately $10^{11}$ cm is formed around it with initial density $10^{-7}$ g/cm$^3$. A further depletion of baryons in the cavity originates from the expansion of the electron-positron-photon ($e^{+}e^{-}\gamma$) plasma formed at the collapse, reaching a density of $10^{-14}$ g/cm$^3$ by the end of the interaction. It is demonstrated here using an analytical model complemented by a hydrodynamical numerical simulation that part of the $e^{+}e^{-}\gamma$ plasma is reflected off the walls of the cavity. The consequent outflow and its observed properties are shown to coincide with the featureless emission occurring in a time interval of duration $t_{rf}$, measured in the rest frame of the source, between $11$ and $20$ s of the GBM observation. Moreover, similar features of the GRB light curve were previously observed in GRB 090926A and GRB 130427A, all belonging to the BdHN I class. This interpretation supports the general conceptual framework presented in \citep{Ruffini2019a} and guarantees that a low baryon density is reached in the cavity, a necessary condition for the operation of the ``{\it inner engine}'' of the GRB presented in an accompanying article \citep{Ruffini2019b}.
\end{abstract}

\section{Introduction}

Soon after the publication of the cosmological redshift value $z=0.42$ for GRB190114C \citep{GCN23695} and the early GBM data obtained by the Fermi-LAT \citep{GCN23709}, \citet{Ruffini2019a} have identified this gamma-ray burst (GRB) as a possible candidate for a binary-driven hypernova of type 1 (BdHN1). Subsequently the division of the GBM data into three distinct episodes \citep{Ruffini2019a} has allowed further  identification of some of the properties of BdHN1-type GRBs. In Episode 1, evidence  was identified for the shock breakout of a supernova giving rise to the entire GRB evolution, see the accompanying article \citep{Li2019}.
BdHNs have as progenitors a carbon-oxygen core (CO$_{core}$) which undergoes a supernova event, thus giving birth to a new neutron star ($\nu$NS) in the presence of a tight binary neutron star (NS) companion. This NS increases its mass by hypercritical accretion until it reaches the critical mass and a black hole (BH) is formed. In this process, a cavity is carved out of the SN ejecta \citep{2016ApJ...833..107B}, see Fig.~\ref{cavity}. Episode 2, observed by the N1, N3, and N4 GBM detectors, includes the dominant ultrarelativistic prompt emission (UPE) phase, and also includes, following the BH formation, the onset of the GeV radiation at a time $t_{rf}=1.9$ s, measured in the rest frame of the source.
Episode 3, which is discussed in this article, addresses the interaction  process between the $e^{+}e^{-}\gamma$ plasma and the cavity. Part of the $e^{+}e^{-}\gamma$ plasma is directed towards the observer and emits the UPE after becoming transparent; another part is reflected by the high density walls of the cavity. Following both the emission of the UPE and the emission from the cavity considered here, a final density of $10^{-14}$ g/cm$^3$ is reached.
This successive series of events leading to a decreasing value of the baryon density is essential to guarantee the further emission of high energy photons, e.g., TeV radiation, and ultrarelativistic cosmic rays \citep{Ruffini2019b}.

The article is structured as follows: in Section 2 we discuss the main features of the light curve of GRB190114C in the context of this scenario, while in Section 3 we provide an analytic model for the interaction between the electron-positron-photon ($e^{+}e^{-}\gamma$) plasma and the baryonic matter in the cavity. In Section 4 we show a hydrodynamical simulation of such a process. In Section 5 we discuss the observed properties of several additional GRBs with a similar structure in their light curves, which support this scenario. In Section 6 we summarize our conclusions.

\section{GRB190114C and the cavity}

At 20:57:02.63 UT on 14 January 2019, GRB 190114C triggered Fermi-GBM, having $T_{90} = 116$ s. The isotropic Fermi-LAT energy is determined as $E_{iso}=1.31\pm 0.19\times 10^{53}$ erg in the energy band of 100 MeV to 100 GeV \citep{2019arXiv190107505W}. The light curve of GRB190114C is shown in Fig.~\ref{grb190114c}, where two major spikes can be distinguished: the first structured spike corresponds to the SN breakout and UPE emission (Episodes 1 and 2) and lasts about 11 s in the rest frame of the source, while the second one (Episode 3) starts after 11 s and lasts for another 9 s. The total energy in the first spike is $95$ percent of $E_{iso}$, while the energy in the second one is 5 percent.

\begin{figure}[th]
\centering
\includegraphics[width=.7\columnwidth]{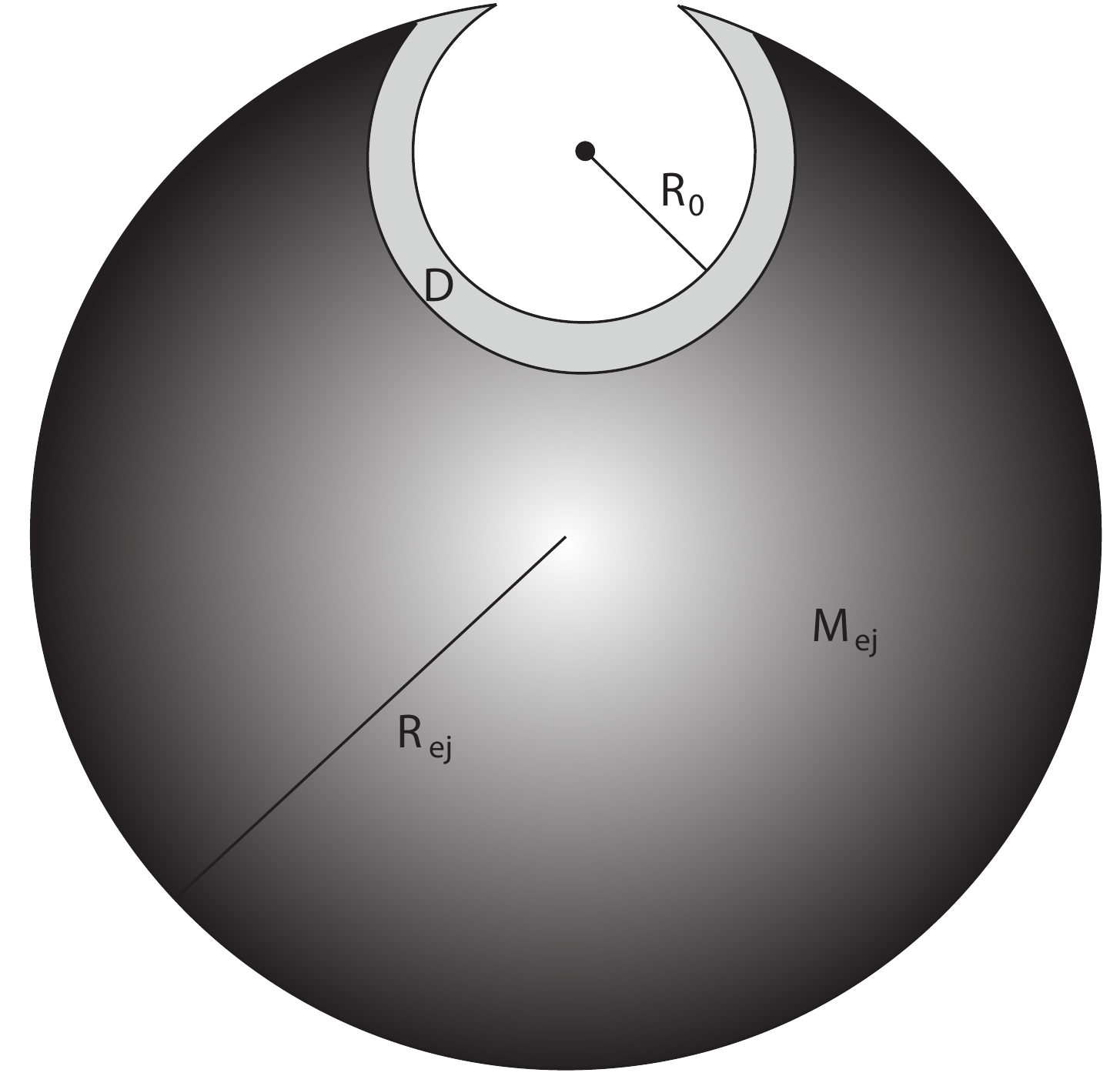}
\caption{Schematic representation of the cavity in the SN\ ejecta surrrounding
the newly formed BH in the BdHN\ scenario.}%
\label{cavity}%
\end{figure}

\begin{figure}[th]
\centering
\includegraphics[width=\columnwidth]{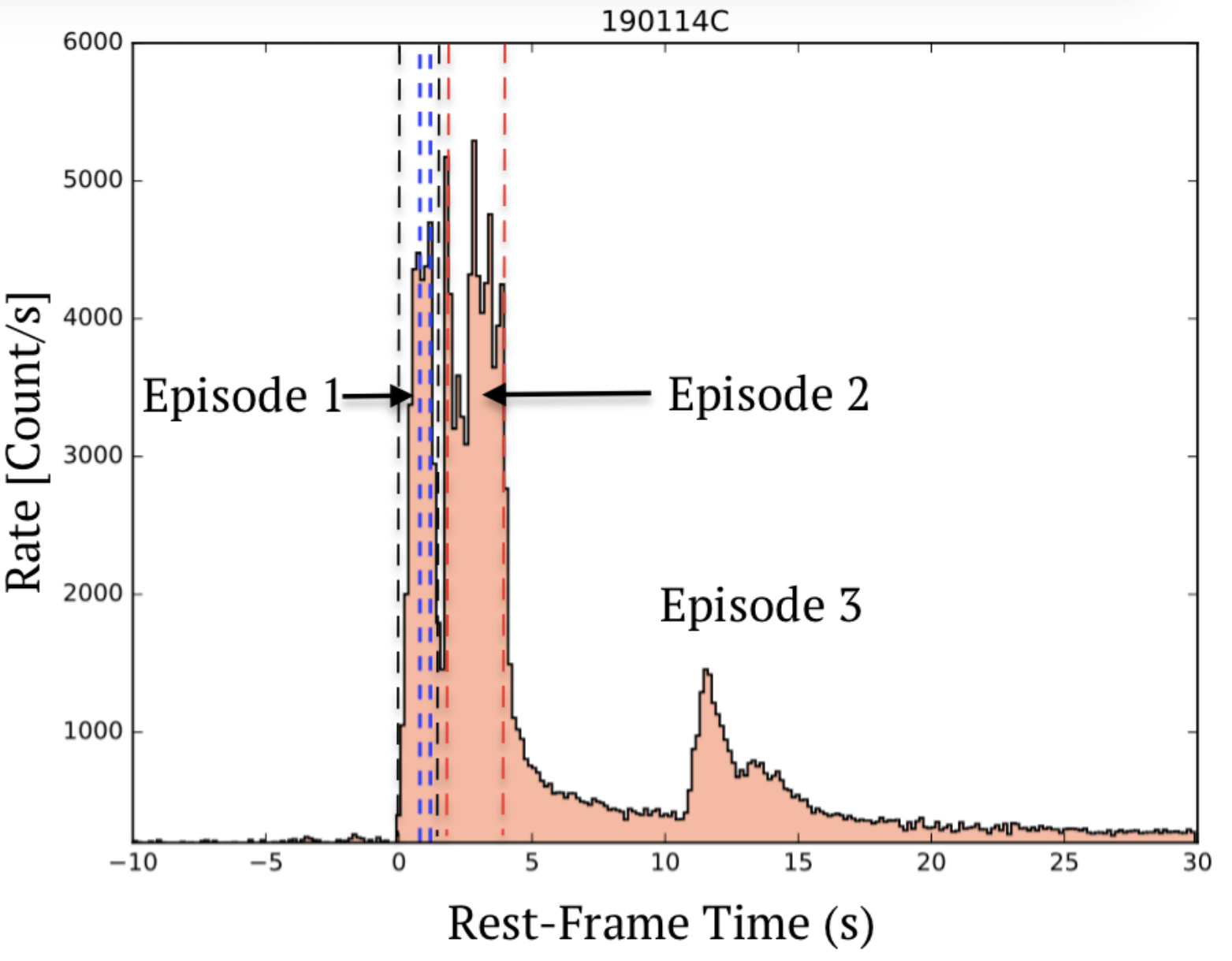}
\caption{The light curve of GRB190114c in the rest frame of the source contains the precursor up to 1.9 s (Episode 1), then UPE emission up to 3.9 s (Episode 2), then the residual emission following the UPE from 3.9 s up to 11 s. The emission from the cavity (Episode 3) occurs at 11 s and continues up to 20 s. The energy of Episode 3 is about 5 percent of the energy in Episode 2.}%
\label{grb190114c}%
\end{figure}

Numerical simulations of the interaction between the SN\ ejecta and the
NS \citep{2016ApJ...833..107B} show that just prior to the BH\ formation the accreting
NS is surrounded by an almost spherical cavity with matter density much lower
than the SN\ ejecta (see Fig.~6 in that paper). These simulations use the smoothed-particle hydrodynamics (SPH) approach, a mesh-free method allowing one to follow the large scale evolution of matter strongly coupled to radiation \citep{2012ApJ...744...52P,2017ApJS..229...27M,2019ApJ...871...14B}. The cavity has an opening
which permits the $e^{+}e^{-}\gamma$ plasma formed around the BH\ to
escape towards the distant observer and emit the
UPE, see Fig.~\ref{cavity}.
Assuming that the $e^{+}e^{-}\gamma$ plasma explodes isotropically, it will also impact
the SN ejecta after propagating in the cavity.

Here we consider the effect of the interaction of the $e^{+}e^{-}\gamma$
plasma with the SN\ ejecta surrounding the cavity, and demonstrate
that such an interaction indeed results in a second spike of GRB emission.
This follows the first main UPE spike produced by the $e^{+}e^{-}\gamma$ 
plasma which
escapes the cavity without interacting with the SN\ ejecta. The second spike
is weaker and delayed with respect to the first one.%

Considering the light curve of GRB190114C shown in Fig.~\ref{grb190114c}, we
argue that the second spike starting at $11$ s and ending at $20$ s can be explained by such an interaction.
Even though the appearance of a cavity is predicted by the mentioned
SPH simulations, we recall that it can also be caused by the expansion of the plasma itself, since the creation of such a lower-density region is a generic feature of strong explosions in a medium (see, e.g., \citep{1994IJBC....4...57V}). Consequently, this model is applicable even in absence of an initial cavity.

\section{Interaction of GRB with the SN\ ejecta near the cavity}

Once the $e^{+}e^{-}\gamma$ plasma is created near the BH, it reaches complete
thermal equilibrium within a very short time, about $10^{-12}$ s
\citep{2007PhRvL..99l5003A,2008AIPC..966..191A}. Since the baryon density inside the cavity is very small,
the plasma produces a blast wave, which expands with acceleration while
transforming its internal energy into kinetic energy \citep{2000A&A...359..855R}. Such an accelerated plasma then impacts the walls of the cavity, which has a radius $R_{0}$ (see Fig.~\ref{cavity}). When the impact occurs, a strong shock wave forms and propagates inside the SN ejecta surrounding the cavity with deceleration until it stalls at a radius $R$. In this way, the kinetic energy of
the $e^{+}e^{-}\gamma$ plasma is deposited as internal energy.

One can determine the thickness of the region $D=R-R_{0}$\ of SN\ ejecta
where this energy is deposited. Assuming that the density of the SN ejecta is
uniform, it can be estimated as%
\begin{equation}
\rho=\frac{3M_{ej}}{4\pi R_{ej}^{3}},\label{ejdens}%
\end{equation}
where $M_{ej}$ is the total mass of the ejecta and $R_{ej}$ is its size at the moment
when the plasma impacts the walls of the cavity. The interaction between
the\ $e^{+}e^{-}\gamma$ plasma and the SN\ ejecta can be considered to be an
inelastic collision. When the $e^{+}e^{-}\gamma$ plasma pulse with total
energy $E$ impacts the ejecta, its energy is transformed into kinetic energy of the
ejecta and internal energy. Assuming that the interaction is adiabatic (e.g., no
energy is lost in radiation during the interaction), energy-momentum
conservation reads%

\begin{align}
E+Mc^{2}  &  =\left(  Mc^{2}+W\right)  \Gamma,\label{encons}\\
\frac{E}{c}  &  =\left(  M+\frac{W}{c^{2}}\right)  \Gamma v,\label{momcons}%
\end{align}
where $\Gamma=\left[  1-\left(  v/c\right)  ^{2}\right]  ^{-1/2}$\ is the
Lorentz factor of the part of ejecta where the energy is deposited, $v$ its
velocity, $W$ the internal energy deposited during the interaction, and $M$ is the
mass of the ejecta affected by the interaction (swept up by the shock wave).

\begin{figure}[th]%
\centering
\includegraphics[width=.8\columnwidth]{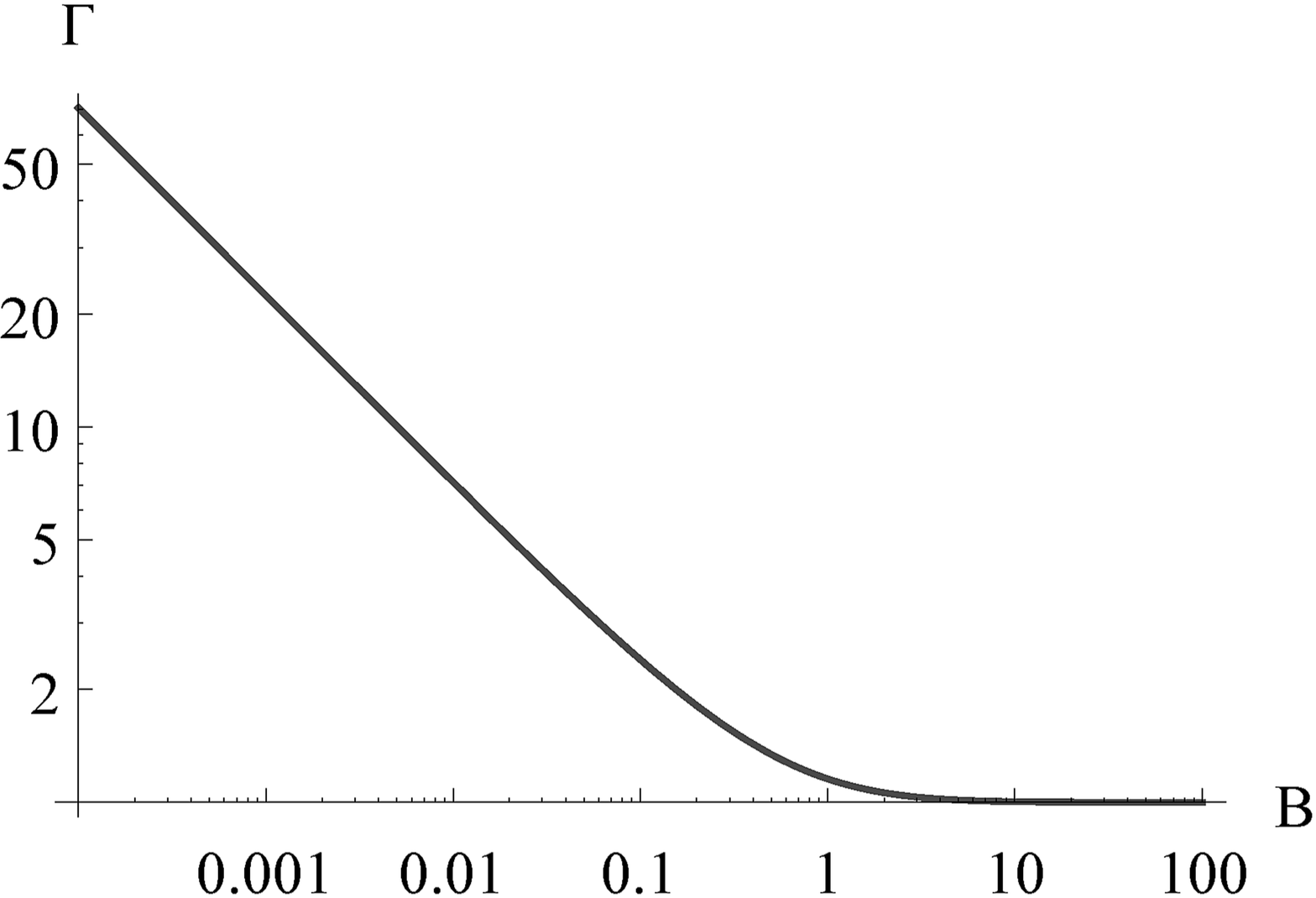}
\caption{Lorentz factor $\Gamma$ of the shock wave generated by the impact of the $e^{+}e^{-}\gamma$
pulse onto the walls of the cavity as a function of the engulfed mass, parametrized by $B=Mc^{2}/E$.}%
\label{gammab}%
\end{figure}

We now introduce the new variables%
\begin{equation}
B=\frac{Mc^{2}}{E},\qquad\omega=\frac{W}{Mc^{2}},\qquad u=\Gamma\frac{v}%
{c}\label{vars}%
\end{equation}
and rewrite the energy-momentum conservation relations as%
\begin{align}
B^{-1} &  =\left(  \omega+1\right)  \sqrt{u^{2}+1}-1,\\
B^{-1} &  =\left(  \omega+1\right)  u.
\end{align}
The solution to this system is \citep{2017A&A...600A.131R}%
\begin{equation}
u=\left(B\sqrt{1+\frac{2}{B}}\right)^{-1},\qquad\frac{W}{E}=\omega B=\frac
{1}{u}-B.\label{Rsolution}%
\end{equation}
Recalling that $\Gamma=\sqrt{1+u^{2}}$, we show $\Gamma(B)$ in Fig.~\ref{gammab}, from which it is clear that the shock wave
stalls when
\begin{equation}
E=Mc^{2}.\label{decmass}%
\end{equation}%
From this figure it is evident that practically all the energy of the $e^{+}%
e^{-}\gamma$\ plasma is deposited as internal energy $W$\ in the region when
the mass-energy of the ejecta is equal to the total energy $E$.

The thickness of this region can be evaluated as follows. We indicate by $M$ the mass engulfed by the shock wave%
\begin{equation}
M=\frac{4\pi}{3}\rho D^{3}.\label{barmass}%
\end{equation}
Inserting (\ref{decmass}) into (\ref{barmass}) allows us to express the thickness as
\begin{equation}
D=\left(  \frac{E}{M_{ej}c^{2}}\right)  ^{1/3}R_{ej}.\label{thickness}%
\end{equation}
For typical parameters%
\begin{equation}
D\simeq2\times10^{11}\left(  \frac{E_{0}}{10^{53}erg}\right)  ^{1/3}\left(
\frac{M_{ej}}{10M_{\odot}}\right)  ^{-1/3}\frac{R_{ej}}{10^{12}cm}%
cm\,.\label{thicknessest}%
\end{equation}
The average temperature in this shocked SN ejecta region can be estimated to be
\begin{align}
T  & =\left(  \frac{3W}{4\pi aR^{2}D}\right)  ^{1/4}\label{tempest}\\
& \simeq12\left(  \frac{W}{10^{53}erg}\right)  ^{1/4}\left(  \frac{D}%
{2\times10^{11}cm}\right)  ^{-3/4}keV.\nonumber
\end{align}
where $a$ is the radiation constant.

After the impact of the $e^{+}e^{-}\gamma$ plasma onto the SN\ ejecta, a
\emph{reflection wave} is generated, which propagates backward into the cavity.
Since the internal energy of the shocked SN\ ejecta is comparable to its rest
mass, the speed of the reflection wave is close to the speed of light.
This reflection wave fills the cavity with baryons and radiation. After the
reflection wave has passed, the cavity expands. Since the cavity is transparent
to radiation, thermal radiation escapes, and produces the second
spike in the emission, analogous to the UPE emission, but observed later and weaker in
intensity. In the next section we present the results of numerical simulations
which support this claim. As the problem under consideration requires the resolution of shock waves on small scales, instead of using the SPH simulations we adopt a different framework based on shock-capturing relativistic hydrodynamics (RHD) \citep[for a discussion of the difference between relativistic SPH and RHD see, e.g.][]{2017rkt..book.....V}.

\section{Numerical simulations}

With the goal to have a deeper insight into the interaction between the
$e^{+}e^{-}\gamma$ plasma and the SN\ ejecta, we have run $2D$ axially symmetric
relativistic hydrodynamic (RHD) simulations using the publicly available PLUTO
code\footnote{\software{PLUTO \citep{2012ApJS..198....7M}}}.
For this approach to be valid, we assume local thermal equilibrium between photons and material particles during the time of the simulation.
\begin{figure}[pth]%
\centering
\includegraphics[width=\columnwidth]{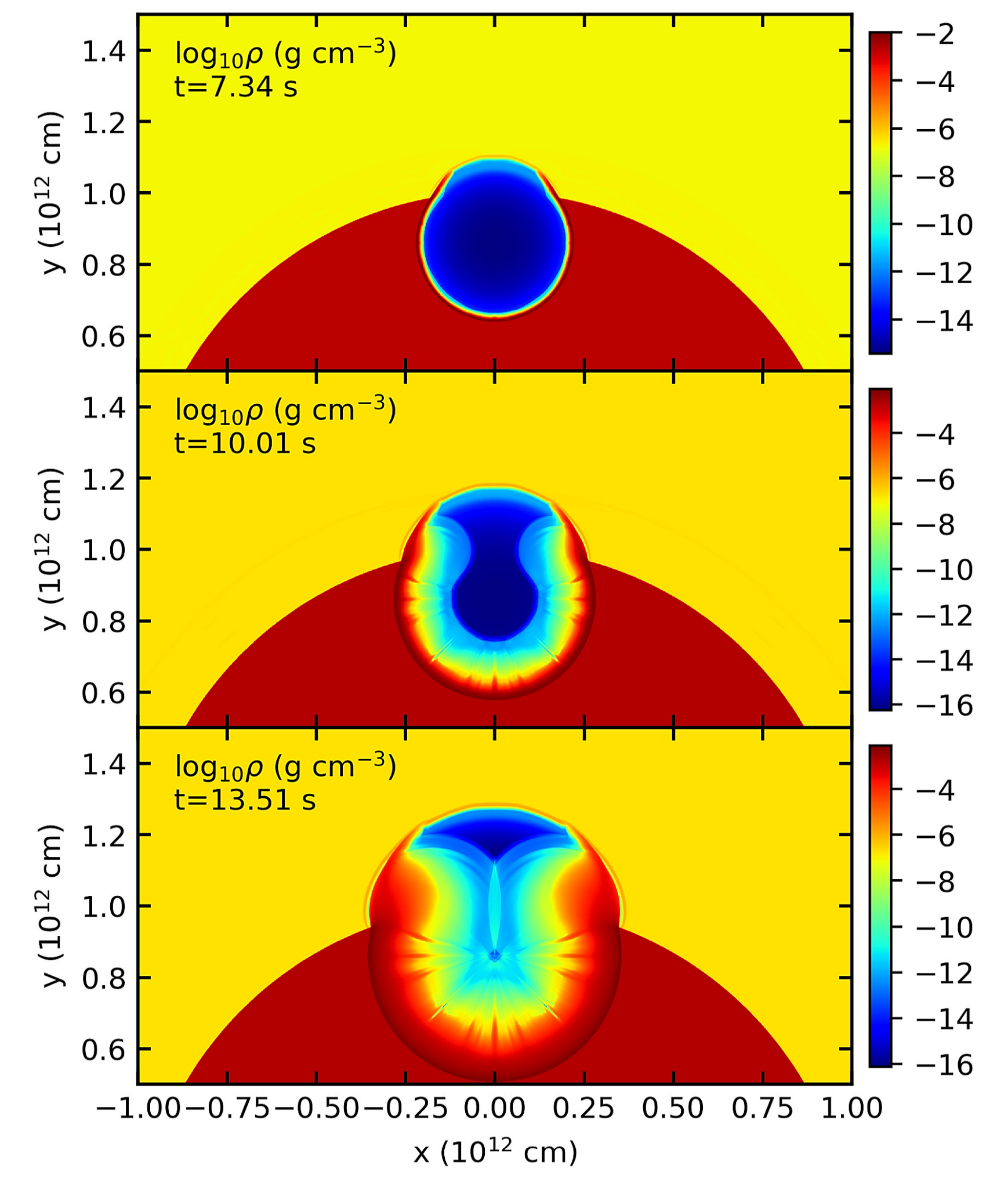}
\caption{Spatial distribution of matter density at different instants of time:\ the time
of impact of the $e^{+}e^{-}\gamma$ plasma onto the cavity walls $t_{imp}=7.3$ s (top), the propagation of the reflection wave
$t_{imp}=10$ s (middle); the reflection wave reaching the origin
$t_{r}=13.5$ s (bottom).}%
\label{density}%
\end{figure}

The consequent equation of state of the resulting
$e^{+}e^{-}\gamma-$baryon plasma is studied in \citep{2018ApJ...852...53R}. In our simulations, we have verified that
applying such equation of state is equivalent to assuming a constant
polytropic index $\gamma_P = 4/3$ in all regions occupied by the plasma
and $\gamma_P=5/3$ in all unshocked regions. We impose this by applying
the so-called TM equation of state as described in \citep{2007MNRAS.378.1118M}.

We assume that a
$e^{+}e^{-}\gamma$ plasma with the total energy $E_{0}=10^{53}$ erg forms in
the center of a spherical cavity, having a radius $R_{cav}=2\times10^{11}$
cm,\ and average density $\rho_{cav}=1.9\times 10^{-7}$ g/cm$^{3}$. The center of the cavity is located at a
distance $1.4\times10^{11}$ cm from the edge of the SN\ ejecta, so that the latter has an opening, as portrayed in Fig.~\ref{cavity}. At this moment the SN\ ejecta, expanding homologously, has
a radius $R_{ej}=10^{12}$ cm.

Since the baryon load in the cavity is small, $B\simeq 10^{-4}$, the
blast wave created by the $e^{+}e^{-}\gamma$ plasma reaches
a bulk Lorentz factor of a few tenths before it impacts the walls of the cavity at $t_{imp}=7.3$
s (see Fig.~\ref{density}). The baryons inside the cavity are swept up by the
expanding plasma, reducing baryon density in the cavity to $\rho_{w}\simeq 10^{-14}$
g/cm$^{3}$. The impact of the $e^{+}e^{-}\gamma$ plasma on the walls of the
cavity produces a shock wave that propagates inside the SN\ ejecta.

\begin{figure}[pth]%
\centering
\includegraphics[width=\columnwidth]{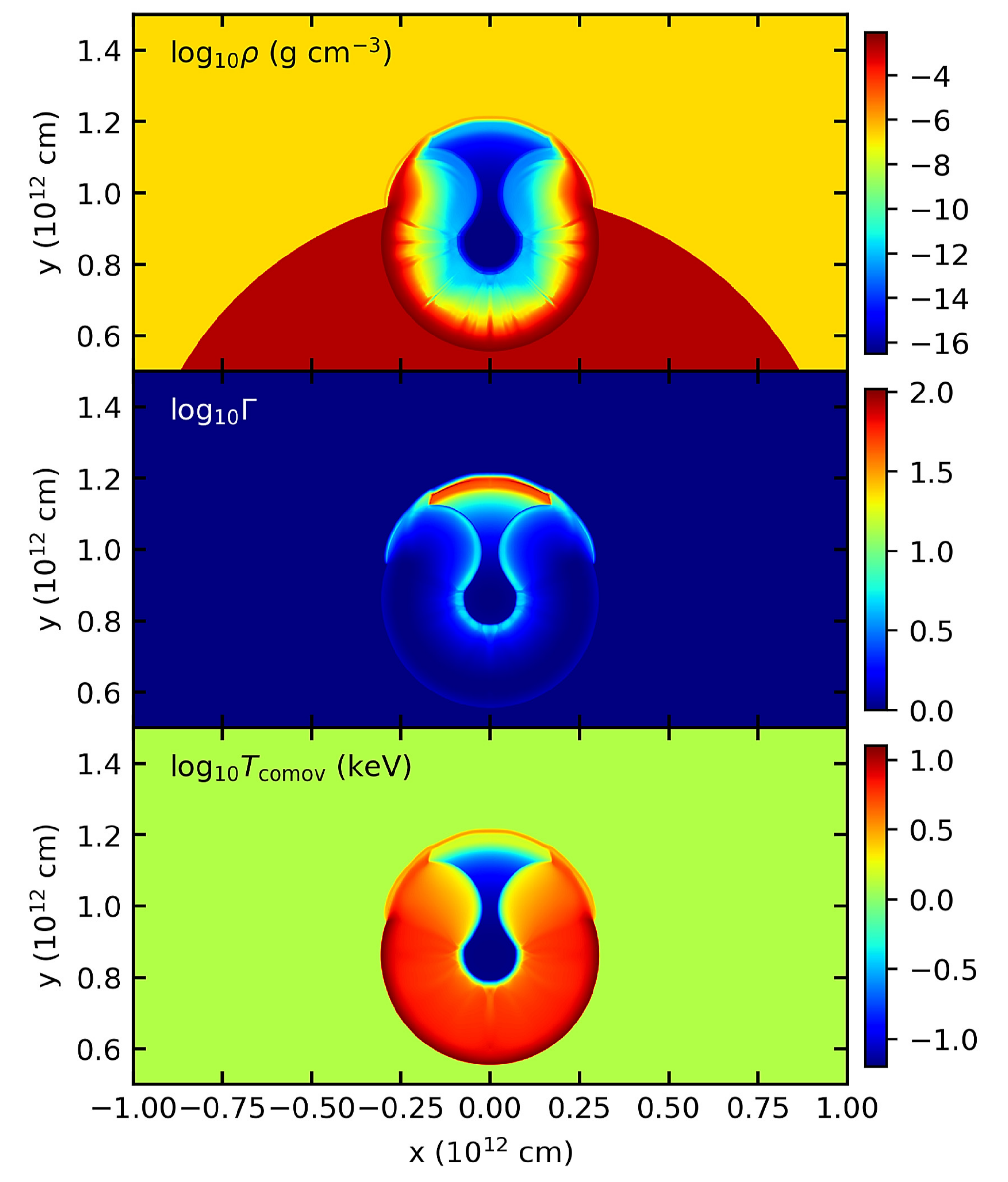}
\caption{Spatial distributions of matter density (top), Lorentz factor (middle) and comoving temperature (bottom) at $t=11$ s, showing the mildly relativistic reflection wave propagating backward in the cavity, as well as the ultrarelativistic $e^{+}e^{-}\gamma$ plasma wave propagating outside the cavity. The shock wave is visible inside the ejecta.}%
\label{2dhydro}%
\end{figure}
After the impact, in agreement with Eq.~(\ref{tempest}), the walls of the cavity are very hot, and their internal energy is radiated inside the cavity. The latter is transparent to radiation, and thus allows free transport of photons through the opening. Inside the walls of the cavity, the radiation drags baryons with it, creating a reflection wave. This wave propagates relativistically backwards into the cavity, leaving a residual
density $\rho_{r}\simeq 10^{-13}$ g/cm$^{3}$ within it, so the total mass of baryons in the
cavity becomes about $M_{r}\simeq 10^{20}$ g.
 The reflection wave eventually reaches
the origin at $t_{r}=13.5$ s. The total energy in the cavity at this point
is $E_{r}=8.3\times10^{51}$ erg, which is only 8.5 percent of the initial
energy $E_{0}$. A large fraction of this energy is emitted towards the observer through the opening, and these estimates
agrees with the energy in the second spike in the light curve in Fig.~\ref{grb190114c}.

\begin{figure}[pth]%
\centering
\includegraphics[width=\columnwidth]{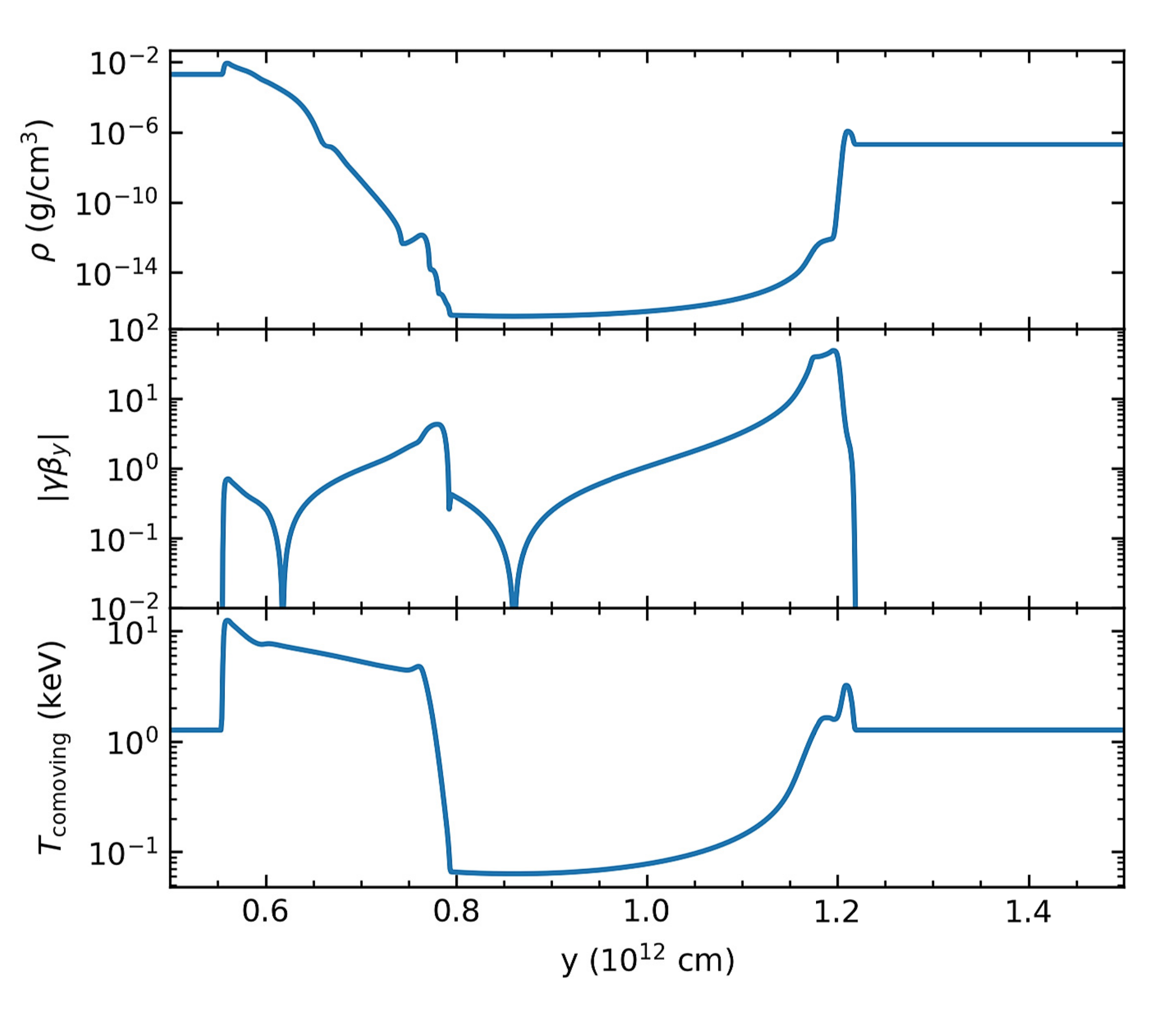}
\caption{Spatial distributions along the axis of symmetry of matter density (top), Lorentz factor (middle) and comoving temperature (bottom) at $t=11$ s. The reflection wave with the maximum Lorentz factor about 4 is clearly visible on the left part of the figure. To the right from the reflection wave the baryons are evacuated by the $e^{+}e^{-}\gamma$ plasma, which has accelerated to the bulk Lorentz factor of about 60.}%
\label{1dhydro}%
\end{figure}
The timescale of emission can be estimated by using the diffusion time formula \citep{2017A&A...600A.131R}
\begin{equation}
\Delta t=3\tau\frac{l}{c},
\label{difftime}
\end{equation}
where $l$ is the thickness of the opaque region and $\tau$ is its optical depth.
Since the density in the cavity wall has a very sharp gradient, while the temperature is nearly uniform, to estimate the duration of the time signal we take the half-thickness of the cavity $l=D/2\simeq 2.5\times 10^{10}$ cm, and the average density $\rho\simeq 10^{-9}$ g/cm$^3$, which gives $\Delta t\simeq 8$ s, in good agreement with the observed duration $9$ s of the second spike.
We also show the distributions of the Lorentz factor, matter density, and comoving temperature at $t=11$ s in Fig.~\ref{2dhydro}. At this point one can clearly see that matter is outflowing from the cavity in the vertical direction, while the oblique shocks propagating from the sides of the cavity opening move towards the vertical axis. The cavity is optically thin and radiation continuously escapes from it to a distant observer.

Naively one expects the spectrum of radiation to be a multicolor blackbody
with a peak energy near $30$ keV. However, our simulations show that
matter in the cavity is moving relativistically towards the observer, reaching a bulk Lorentz factor $\gamma\simeq 3$ (see Fig.~\ref{1dhydro}). Due to the non-negligible optical depth, the radiation generated by the walls of the
cavity will experience inverse Compton scattering on this relativistically moving matter,
Doppler-shifting the peak energy towards higher values. Therefore, we expect
the observed spectrum to be similar to a Comptonized black body \citep{2013MNRAS.436L..54A} peaked
at $E_p\sim 3\gamma^2 kT\simeq 200-300$ keV. This qualitatively agrees with the observed spectrum of the second
spike of GRB190114C. The time-integrated spectrum has a very soft power law index $\alpha=-1.6$ and a peak energy $E_p=252$ keV. The time-resolved spectrum shows spectral evolution towards softening of the spectrum, in agreement with the cooling of the expanding walls of the cavity.

\vspace{.6in}

\section{Additional cases with Episode 3}

\begin{figure}[pbh]%
\centering
\includegraphics[width=\columnwidth]{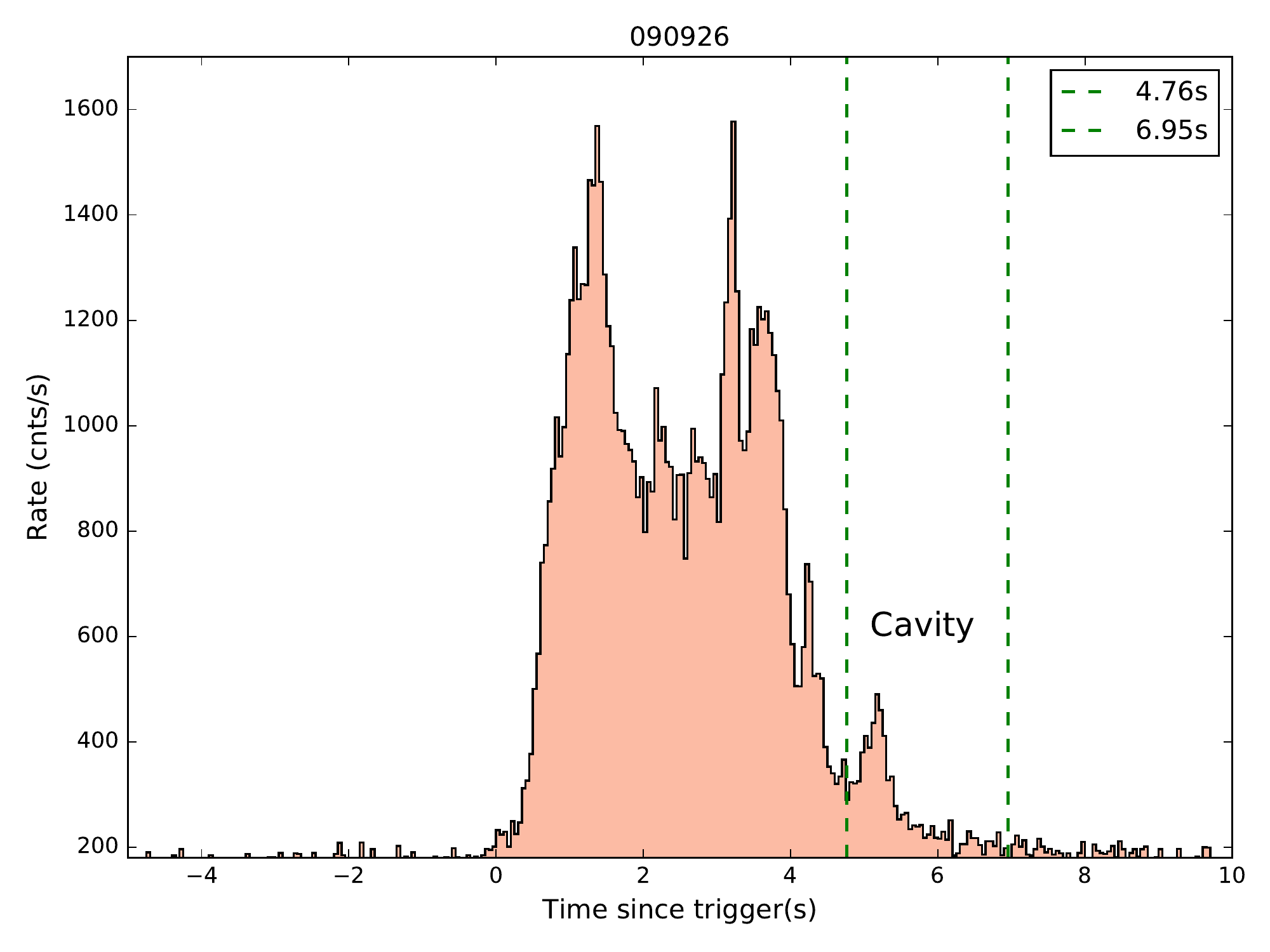}
\includegraphics[width=\columnwidth]{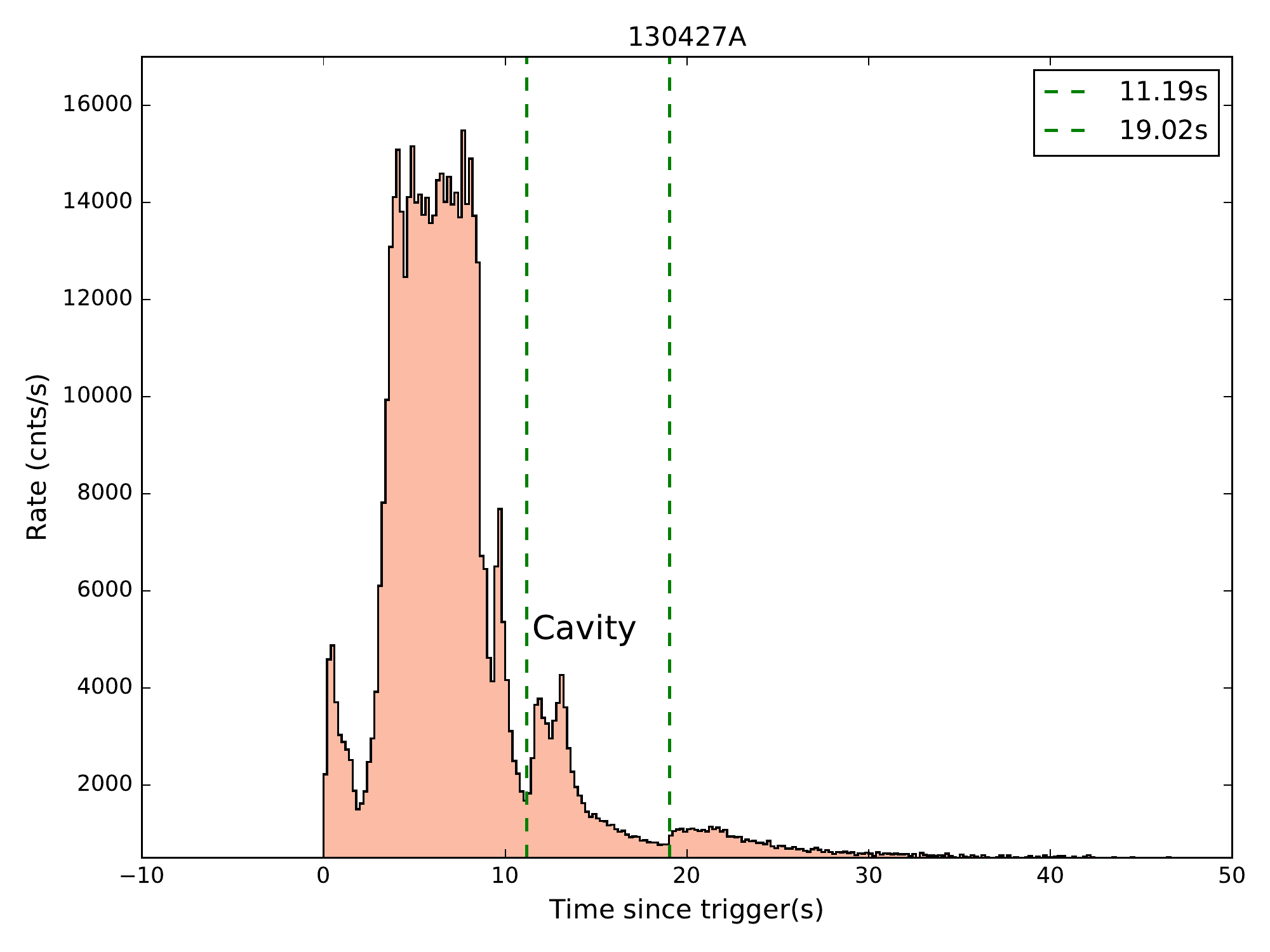}
\caption{Light curves of GRB 090926A (upper panel) and GRB 130427A (lower panel), in the source rest frame. The structure of the light curves as well as the spectral analysis allows to identify the Episode 3 in these GRBs, for details see \citep{Li2019a}. Data from Fermi GBM satellite.}%
\label{2GRBs}%
\end{figure}

A similar light curve structure, i.e. with an Episode 3 well separated from the UPE emission, has been identified in several GRBs. In particular, that is the case of GRB 090926A and GRB 130427A. All of these sources belong to the BdHN I class, as they all have an isotropic energy above $10^{52}$ erg: for GRB 090926A $E_{iso}=1.89\times 10^{54}$ erg, while for GRB 130427A $E_{iso}=1.06\times 10^{54}$ erg. Furthermore, all of them have a featureless emission occurring with a delay with respect to the main emission of Episode 2 of $t_{rf}\simeq5$ sec for GRB 090926A $t_{rf}\simeq11$ sec for GRB 130427A, see Fig. \ref{2GRBs}.
The spectra of Episode 3 in both cases are well fitted by cut-off power law model. For details about the light curve and spectra of these GRBs see \citep{Li2019a}. These observations support the interpretation of these three GRB light curves within the BdHN model, and are consistent with the occurrence of reflection waves with properties similar to the one presented above.

We recall that the systems analyzed in this article have as progenitors a binary system composed of CO$_{core}$ undergoing SN explosion, and consequently hypercritically accreting into a companion NS. As shown above, three of these systems have been clearly identified in GRBs 090926A, 130427A and 190114C, see Figs. \ref{grb190114c} and \ref{2GRBs} and \citep{Li2019a}.

One of the most interesting aspects of the evolution of these binary systems is that they evolve in a succession of GRB emissions. It is so that the system treated in this article describing long GRBs, with progenitor composed of a CO core and NS may produce as an outcome a binary system composed of a NS and BH which evolves after $\sim10^4$ years in a new merging, giving origin to a short GRB \citep{2015PhRvL.115w1102F}.

Such NS-BH systems were indeed among the first to be considered as progenitors of short GRBs in the pioneering works of \citet{1991AcA....41..257P} and \citet{1999Sci...284..115V}. \citet{1999Sci...284..115V} considered the possibility of toroidal magnetic field structure and cavities in such a system. This work was further developed in great details in \citep{2003ApJ...584..937V}. This field has received a revival thanks to numerical simulations of the toroidal magnetic field created in the NS-BH and NS-NS merging \cite{2011ApJ...732L...6R,2017MNRAS.469L..31N} and references therein. It is interesting that such systems should be quite common \citep{2015PhRvL.115w1102F}. They may be related to the problematic of fast radio bursts and to ultra-short GRBs. It is clear that the cavity considered in SN accretion on a NS leading to formation of BH, and the cavity originating from the NS-BH merging address different physical problematics.

\section{Conclusions}

We have presented an interpretation of the light curve of GRB190114C as the result of BH formation in the BdHN scenario. In this scenario a massive CO$_{core}$, which forms a binary system with a NS, explodes as a supernova. The NS undergoes gravitational collapse due to hypercritical accretion, creating in the process a low-density cavity in the SN ejecta. In this picture the first spike (UPE) originates from an $e^{+}e^{-}\gamma$ plasma formed around the black hole, while the second spike originates from the reflection wave produced by the impact of the plasma on the walls of the cavity. Our $2D$ relativistic hydrodynamic simulations support such a picture and match the energetics and timing characteristics of the signal. Spectral characteristics and their time evolution are in qualitative agreement with the observations. Similar features of GRB light curves have been recently observed in GRB 090926A and GRB 130427A, all belonging to the BdHN 1 class. This is consistent with a
rather generic occurrence of reflection waves such as the one described
in this work.

\acknowledgments We thank Carlo Luciano Bianco and She-Sheng Xue for useful discussions and Dr. Li Liang for supplying figures 2 and 7. We are also grateful to the Referee for his/her important remarks which allowed us to focus more clearly on our key results.

\end{document}